\documentclass[10pt]{article}
\usepackage[T1]{fontenc}
\usepackage[latin9]{inputenc}

\makeatletter

\let\SF@@footnote\footnote
\def\footnote{\ifx\protect\@typeset@protect
    \expandafter\SF@@footnote
  \else
    \expandafter\SF@gobble@opt
  \fi
}
\expandafter\def\csname SF@gobble@opt \endcsname{\@ifnextchar[
  \SF@gobble@twobracket
  \@gobble
}
\edef\SF@gobble@opt{\noexpand\protect
  \expandafter\noexpand\csname SF@gobble@opt \endcsname}
\def\SF@gobble@twobracket[#1]#2{}

\providecommand{\tabularnewline}{\\}

\usepackage{M12}
\makeatother

\begin{document}
\setpreprint{QGaSLAB-13-05, \href{http://arxiv.org/abs/1306.5106}{arXiv:1306.5106}} 
\settitlesize{14.6pt}

\title{The $AdS_{3}\times S^{3}\times S^{3}\times S^{1}$ Hern\'{a}ndez--L\'{o}pez
Phases: a Semiclassical Derivation}

\author{Michael C. Abbott  \address{Department of Mathematics, University
of Cape Town, South Africa.\\
michael.abbott@uct.ac.za} }

\date{21 June 2013 (v1)}
\maketitle
\begin{abstract}
This note calculates the Hern\'{a}ndez--L\'{o}pez phases for strings
in $AdS_{3}\times S^{3}\times S^{3}\times S^{1}$ by semiclassical
methods using the $d(2,1;\alpha)^{2}$ algebraic curve. By working
at general $\alpha$ we include modes absent from previous semiclassical
calculations of this phase in $AdS_{3}\times S^{3}\times T^{4}$,
and in particular can study the scattering of particles of different
mass. By carefully re-deriving the semiclassical formula we clarify
some issues of antisymmetrisation, cutoffs and surface terms which
could safely be ignored in $AdS_{5}\times S^{5}$, and some issues
about the terms $c_{1,s}$ which were absent there. As a result we
see agreement with the recently calculated all-loop dressing phase
in the $AdS_{3}\times S^{3}\times T^{4}$ case, and exactly $1/2$
this in the general case $AdS_{3}\times S^{3}\times S^{3}\times S^{1}$,
for any $\alpha$ and any (light) polarisations. 
\end{abstract}

\section{Introduction}

Our understanding of integrability in the AdS/CFT correspondence between
$\ads5\times\s5$ and $\mathcal{N}=4$ SYM is largely contained in
the exact S-matrix and its associated Bethe equations \cite{Beisert:2010jr}.
Much of this is fixed by symmetry of the spin-chain, but some remaining
freedom is filled in by the celebrated Beisert--Eden--Staudacher dressing
phase \cite{Beisert:2006ez}. The dressing phase was first introduced
 by Arutyunov, Frolov and Staudacher (AFS) \cite{Arutyunov:2004vx}
for classical strings (large $\lambda$), and the extension of this
to one loop by Hern\'{a}ndez and L\'{o}pez \cite{Hernandez:2006tk}
(see also \cite{Beisert:2005cw,Freyhult:2006vr,Beisert:2006ib}) was
an important step towards the all-$\lambda$ phase, conventionally
$(\sigma_{\mathrm{BES}})^{2}$. This phase was later observed at small
$\lambda$ (four loops) by \cite{Beisert:2007hz}.

The first example of AdS/CFT with less supersymmetry to be well-explored,
$\ads4\times\cp3$ and ABJM theory \cite{Aharony:2008ug,Klose:2010ki},
turns out to have an almost identical dressing phase: one power of
$\sigma_{\mathrm{BES}}$ \cite{Gromov:2008qe,Mauri:2013vd}. The subject
of this note is IIB strings in the R-R background $\ads3\times\s3\times\s3\times\s1$,
which along with the related background $AdS_{3}\times S^{3}\times T^{4}$
has been the subject of much recent work in integrability \cite{David:2008yk,Babichenko:2009dk,Zarembo:2010yz,David:2010yg,OhlssonSax:2011ms,Forini:2012bb,Rughoonauth:2012qd,Sundin:2012gc,Cagnazzo:2012se,Sax:2012jv,Ahn:2012hw,Borsato:2012ud,Abbott:2012dd,Beccaria:2012kb,Borsato:2012ss,Beccaria:2012pm,Sundin:2013ypa,Borsato:2013qpa,Borsato:2013hoa}.%
\footnote{There is related work on deformations \cite{Mohammedi:2008vd,Orlando:2010yh,Kawaguchi:2011mz,Kawaguchi:2011pf,Azeyanagi:2012zd,Orlando:2012hu},
quotients \cite{David:2011iy,David:2012aq}, S-matrices \cite{Hoare:2011fj,Bianchi:2013nra,Engelund:2013fja},
and mixed NS-NS backgrounds \cite{Hoare:2013pma,Hoare:2013ida}.%
} It has been discovered that while the AFS phase is modified in a
quite simple way, the HL phase is different to that seen before. This
was seen in comparisons of one-loop energy corrections \cite{Abbott:2012dd},
in direct semiclassical calculations of the phase by \cite{David:2010yg,Beccaria:2012kb}
(for the $T^{4}$ case), and in near-BMN and near-flat-space calculations
\cite{Sundin:2013ypa}. 

There is now a proposal for an all-loop dressing phase for the $T^{4}$
case \cite{Borsato:2013hoa}. When this is expanded in the coupling,
the coefficients $c_{r,s}$ in the HL terms are slightly different
to those written down by \cite{Beccaria:2012kb}. 

The goal of this note is to provide a clear semiclassical derivation
of the phases for the $AdS_{3}\times S^{3}\times S^{3}\times S^{1}$
case, following the method of \cite{Chen:2007vs} but being careful
to keep all surface terms dropped there. We find coefficients exactly
$1/2$ those found by \cite{Borsato:2013hoa}, independent of $\alpha$,
and unaffected by whether the scattering involves light particles
of the same or different mass.%
\footnote{The two $S^{3}$ factors are of radius $R/\negmedspace\cos\phi$ and
$R/\negmedspace\sin\phi$, where $R$ is the $AdS_{3}$ radius and
$\phi$ (or often $\alpha=\cos^{2}\phi$) is a free parameter of the
theory. As a result of this there are modes of mass $1$, $\cos^{2}\phi$,
$\sin^{2}\phi$ and $0$, approximately corresponding to the factors
$\ads3\times\s3\times\s3\times\s1$. We are mostly concerned with
the light modes, $m=\cos^{2}\phi,\sin^{2}\phi$, as these are fundamental
in the description of \cite{Borsato:2012ud,Borsato:2012ss}. In the
limit $\phi\to0$ the second sphere decompactifies, taking us to $\ads3\times\s3\times T^{4}$
at least in terms of the bosonic geometry. %
} It is easy to go to the $T^{4}$ case by omitting the modes corresponding
to one $S^{3}$ factor, in which case the agreement with \cite{Borsato:2013hoa}
is perfect. 

Antisymmetry of the phase under the exchange of the two physical particles
emerges naturally. In fact there is both a symmetric and an antisymmetric
term in the general result (\ref{eq:my-IXY-IX-organisation}), and
in the magnon regime $p\sim1$ the former cancels against surface
terms. If however we consider the regime $p\ll1$, then the instead
the symmetric and antisymmetric terms cancel against each other to
remove the terms linear in $p$ (or in $x$).

\subsection*{Outline}

Section \ref{sec:Deriving} looks at one-loop semiclassical phases
in general, and section \ref{sec:Applying} applies this to magnons
in $AdS_{3}\times S^{3}\times S^{3}\times S^{1}$ (including left-right
\S\,\ref{sub:The-left-right-phase} and unequal mass \S\,\ref{sub:Particles-of-unequal-mass}).
Section \ref{sec:Conclusions} concludes. Appendix \ref{sec:Algebraic-Curve-Setup}
sets up the derivation of the classical phase shifts which are the
input for this calculation.

\section{Deriving the One-Loop Phase\label{sec:Deriving}}

The same classical data used to find the one-loop correction to the
energy \cite{Dashen:1975hd} can also be used to find the one-loop
correction to the scattering phase. This idea, originally from \cite{Faddeev:1977rm},
was used by \cite{Chen:2007vs} to recover the Hern\'{a}ndez--L\'{o}pez
phase in $\ads5\times\s5$. (A similar idea was also developed by
the slightly earlier paper \cite{Gromov:2007cd}.)

The one-loop correction to the energy is given by a sum of frequencies
$\omega_{n}$ over all mode numbers $n$, and polarisations $r$.
This can often be written as an integral over momentum $k$, and the
necessary Jacobians for this are derivatives of the classical phase
shifts $\delta_{r}(k)$ experienced by modes of the soliton: 
\begin{align}
\Delta E & =\sum_{r}(-1)^{F_{r}}\Big[\sum_{n}\frac{1}{2}\omega_{n}^{r}-\sum_{n}\frac{1}{2}\omega_{n\:\mathrm{vac}}^{r}\Big]\label{eq:dE-simple}\\
 & =\sum_{r}(-1)^{F_{r}}\frac{1}{4\pi}\int dk\:\frac{\partial\delta_{r}(k)}{\partial k}\:\omega^{r}(k).\nonumber 
\end{align}
Rather than working in momentum space we will work in the spectral
plane, in which case we obtain a similar integral with $\partial_{z}\delta_{r}(z)\:\Omega^{r}(z)$.
For $\ads3$ this kind of energy correction has been discussed by
\cite{David:2010yg,Forini:2012bb,Abbott:2012dd,Beccaria:2012kb,Beccaria:2012pm},
starting from a variety of classical systems. 

The one-loop correction to the scattering phase is given by \cite{Chen:2007vs}
in terms of the phase shifts $\delta_{r}(k,p)$ (for a mode $k$ in
the background of a single magnon $p$) as follows:
\begin{equation}
\Delta\Theta(p_{1},p_{2})=\frac{1}{2\pi}\int dk\sum_{r}(-1)^{F_{r}}\left[\frac{\partial\delta_{r}(k,p_{1})}{\partial k}\delta_{r}(k,p_{2})\right].\label{eq:dTheta-one-sided-CDLM}
\end{equation}
However na\"{i}vely applying this formula to the present case leads
to a correction which is not antisymmetric in $x\leftrightarrow y$.
This was cured in \cite{David:2010yg} by explicitly antisymmetrising,
noting that this amounts to keeping one of the surface terms dropped
by \cite{Chen:2007vs}. Various approaches to this issue were explored
in \cite{Beccaria:2012kb}. 

To derive (\ref{eq:dTheta-one-sided-CDLM}), Chen, Dorey and Lima
Matos \cite{Chen:2007vs} studied a classical scattering state $\varphi_{1,2}(x,t)$
of two magnons in a box of size $L$, and calculated $1/L$ corrections
to the one-loop energy correction (\ref{eq:dE-simple}). Their notation
is that quantum corrections $1/g$ are written\newcommand{\Eclass}{E_\mathrm{cl.}} \newcommand{\Eclassone}{E_{\mathrm{cl.}1}} \newcommand{\Eclasstwo}{E_{\mathrm{cl.}2}} \newcommand{\Eclassi}{E_{\mathrm{cl.}i}}
\begin{equation}
E(p)=g\Eclass(p)+\Delta E(p)+\smash{\bigodiv{g}}\label{eq:CDLM-notation-g}
\end{equation}
and corrections in $1/L$ are written 
\[
p=p^{(0)}+\frac{1}{L}p^{(1)}+\smash{\bigodiv{L^{2}}},\qquad\mbox{with }p^{(0)}=\frac{2\pi}{L}n.
\]
Consider two physical particles with mode numbers $n_{1}$ and $n_{2}$,
and for short write $p_{n_{1}}=p_{1}$ etc. They have polarisations
$r_{1}$, $r_{2}$, for which we similarly write $\Theta_{r_{1}r_{2}}=g\Theta_{\mathrm{cl.}12}+\Delta\Theta_{12}+\ldots$.
To find the one-loop phase $\Delta\Theta_{12}$, we calculate the
one-loop order-$1/L$ correction $\Delta E^{(1)}$ in two different
ways:
\begin{itemize}
\item The first calculation is about the quantisation conditions for the
momenta. Demanding periodicity of the solution in the box $L$, we
obtain
\[
p_{1}L=2\pi n_{1}+\Theta_{12}(p_{1},p_{2}),\qquad p_{2}L=2\pi n_{2}+\Theta_{21}(p_{2},p_{1}).
\]
The $1/L$ term is 
\[
p_{1}^{(1)}=\Theta_{12}(p_{1}^{(0)},p_{2}^{(0)})=g\Theta_{\mathrm{cl.}12}(p_{1}^{(0)},p_{2}^{(0)})+\Delta\Theta_{12}(p_{1}^{(0)},p_{2}^{(0)})+\bigodiv{g}
\]
and similarly for $p_{2}$. Since scattering is elastic, the energy
of a two-particle state is $E=E_{1}(p_{1})+E_{2}(p_{2})$. These functions
receive $1/g$ corrections but not $1/L$ corrections, so the only
such correction comes from the shifts of the momenta: 
\[
E^{(1)}=\sum_{i=1}^{2}\frac{dE_{i}(p_{i}^{(0)})}{dp_{i}}p_{i}^{(1)}=\sum_{i}\left[g\Eclassi'(p_{i}^{(0)})\: p_{i}^{(1)}+\Delta E_{i}'(p_{i}^{(0)})\: p_{i}^{(1)}\right]+\bigodiv{g}.
\]
Using $p_{i}^{(1)}$ above, the one-loop part of this is
\begin{align}
\Delta E^{(1)} & =g\Eclassone'(p_{1}^{(0)})\:\Delta\Theta_{12}(p_{1}^{(0)},p_{2}^{(0)})+g\Eclasstwo'(p_{2}^{(0)})\:\Delta\Theta_{21}(p_{2}^{(0)},p_{1}^{(0)})\nonumber \\
 & \quad+\Delta E_{1}'(p_{1}^{(0)})\: g\Theta_{\mathrm{cl.}12}(p_{1}^{(0)},p_{2}^{(0)})+\Delta E_{2}'(p_{2}^{(0)})\: g\Theta_{\mathrm{cl.}21}(p_{2}^{(0)},p_{1}^{(0)}).\label{eq:result-of-path-1-1}
\end{align}
The second line is absent in \cite{Chen:2007vs}. (However for magnons
in $AdS_{5}$ we have $\Delta E(p)=0$.)
\item The second calculation is a zero-point-energy mode sum, like (\ref{eq:dE-simple}).
The frequencies here are those of a third classical particle (mode
$n$, momentum $k_{n}$) on top of the same two physical particles.
We must include in $\omega_{n}$ the shifts in the energies of the
other two particles due to the extra phase shift, which we write as
$g\Theta_{\mathrm{cl}.\, r\,1}=\delta_{r1}=-\delta_{1r}$. Summing
the $1/L$ term of $\omega_{n}$ then gives 
\begin{equation}
\negthickspace\negthickspace\negthickspace\Delta E_{r}^{(1)}=\frac{g}{2}\sum_{n}\left\{ \Eclassone'(p_{1})\left[\delta_{r1}(k_{n},p_{1})-\delta_{r1}(k_{n}^{(0)},p_{1})\right]+\Eclasstwo'(p_{2})\left[\delta_{r2}(k_{n},p_{2})-\delta_{r2}(k_{n}^{(0)},p_{2})\right]\right\} \label{eq:result-of-path-2-1}
\end{equation}
as in \cite{Chen:2007vs}. The subscript $r$ is to indicate that
we still need to sum over all polarizations. 
\end{itemize}
Equating (\ref{eq:result-of-path-1-1}) and (\ref{eq:result-of-path-2-1}),
note that we must assume $\Delta\Theta_{21}=-\Delta\Theta_{12}$ in
order to get an answer at all. Then we get the following one-loop
correction to the scattering phase:
\begin{align}
\negthickspace\negthickspace\negthickspace\Delta\Theta_{12}(p_{1},p_{2}) & =\sum_{r}(-1)^{F_{r}}\frac{1}{2}\sum_{n}\Bigg\{\frac{\Eclassone'(p_{1})}{\Eclassone'(p_{1})-\Eclasstwo'(p_{2})}\left[\delta_{r1}(k_{n},p_{1})-\delta_{r1}(k_{n}^{(0)},p_{1})\right]\displaybreak[0]\nonumber \\
 & \qquad\qquad\qquad\quad+\frac{\Eclasstwo'(p_{2})}{\Eclassone'(p_{1})-\Eclasstwo'(p_{2})}\left[\delta_{r2}(k_{n},p_{2})-\delta_{r2}(k_{n}^{(0)},p_{2})\right]\Bigg\}\nonumber \\
 & \qquad\qquad-\frac{\Delta E_{1}'(p_{1})-\Delta E_{2}'(p_{2})}{g\Eclassone'(p_{1})-g\Eclasstwo'(p_{2})}\: g\Theta_{\mathrm{cl.}12}(p_{1},p_{2}).\label{eq:DeltaTheta-sum-n}
\end{align}
To convert the sums over $n$ into integrals, the relevant Jacobians
are 
\[
2\pi\frac{\partial n}{\partial k}=L+\frac{\partial\delta_{r1}(k,p_{1})}{\partial k}+\frac{\partial\delta_{r2}(k,p_{2})}{\partial k},\qquad2\pi\frac{\partial n}{\partial k^{(0)}}=L
\]
leading to integrals of the form\newcommand{\newI}{I} 
\[
\newI_{XY}=\frac{1}{8\pi}\int dk\frac{\partial\delta_{r1}(k,p_{X})}{\partial k}\delta_{r2}(k,p_{Y}),\qquad I_{X}=I_{XX}
\]
where we now write $p_{1}=p_{X}$ and $p_{2}=p_{Y}$. Clearly $I_{X}$
is a surface term. It will be convenient to organise the terms like
this: 
\begin{equation}
\Delta\Theta_{12,r}(p_{X},p_{Y})=(\newI_{YX}-\newI_{XY})+(I_{X}-I_{Y})+\frac{\Eclassone'(p_{X})+\Eclasstwo'(p_{Y})}{\Eclassone'(p_{X})-\Eclasstwo'(p_{Y})}\left[(I_{XY}+I_{YX})+(I_{X}+I_{Y})\vphantom{\frac{1}{1}}\right].\label{eq:my-IXY-IX-organisation}
\end{equation}
The total is then $\Delta\Theta_{r_{1}r_{2}}=\sum_{r}(-1)^{F_{r}}\Delta\Theta_{12,r}+(\text{term in }\Delta E'\Theta_{\mathrm{cl.}})$.
We treat the last term here separately below, and focus mostly on
(\ref{eq:my-IXY-IX-organisation}).

\section{$AdS_{3}$ Magnon Scattering\label{sec:Applying}}

The phase shifts we need are those for a mode in the background of
a single giant magnon. These were calculated in \cite{Abbott:2012dd}
for the purpose of finding energy corrections (\ref{eq:dE-simple}),
and they are written in terms of the spectral parameter as 
\begin{equation}
\delta_{r}(z,\Xpm)=2\pi n_{r}(z)_{\mathrm{mag}(\Xpm)}-2\pi n_{r}(z)_{\mathrm{vac}}.\label{eq:delta-is-two-pi-n-with-vac}
\end{equation}
These are shown in table \ref{tab:List-of-phase-shifts}; see appendix
\ref{sec:Algebraic-Curve-Setup} for details. 

\begin{table}
\centering %
\begin{tabular}{cc|cc|cc|cc}
 &  & \multicolumn{2}{c|}{\textbf{$\ads3\times\s3\times\s3$}} & \multicolumn{2}{c}{\textbf{$\ads3\times\s3$} } &  & \tabularnewline
$r$ & $m_{r}$ & \multicolumn{2}{c|}{$\delta_{r,3}(z)$} & \multicolumn{2}{c|}{$\delta_{r,(\tilde{1}\tilde{4})}(z)$} & $r=(i,j)$ & $m_{r}$\tabularnewline
\hline 
$0$, $\bar{0}$ & 0 &  & $\vphantom{\tfrac{1}{z}}$ &  &  &  & 0\tabularnewline
$0f$, $\bar{0}f$ & 0 &  & $\vphantom{\tfrac{1}{z}}$ &  &  &  & 0\tabularnewline
\cline{1-4} 
$1$, $\bar{1}$ & $\sin^{2}\phi$ & $0$, & $0$ &  &  &  & 0\tabularnewline
$1f$, $\bar{1}f$ & $\sin^{2}\phi$ & $-G(z)$, & $G(\tfrac{1}{z})$ &  &  &  & 0\tabularnewline
\hline 
\textbf{$3$}, $\bar{3}$ & $\cos^{2}\phi$ & \textbf{$2G(z)$,} & $-2G(\tfrac{1}{z})$ & \textbf{$2G(z)$,} & $-2G(\tfrac{1}{z})$ & \textbf{$(\tilde{1},\tilde{4})$}, $(\tilde{2},\tilde{3})$ & 1\tabularnewline
$3f$, $\bar{3}f$ & $\cos^{2}\phi$ & $G(z)$, & $-G(\tfrac{1}{z})$ & $G(z)$, & $-G(\tfrac{1}{z})$ & $(\hat{1},\tilde{4})$, $(\hat{2},\tilde{3})$ & 1\tabularnewline
\cline{1-4} 
$4$, $\bar{4}$ & 1 & $0$, & $0$ & $0$, & $0$ & $(\hat{1},\hat{4})$, $(\hat{2},\hat{3})$ & 1\tabularnewline
$4f$, $\bar{4}f$ & 1 & $G(z)$, & $-G(\tfrac{1}{z})$ & $G(z)$, & $-G(\tfrac{1}{z})$ & $(\tilde{1},\hat{4})$, $(\tilde{2},\hat{3})$ & 1\tabularnewline
\hline 
\end{tabular}\caption{Phase shifts in modes of the $AdS_{3}\times S^{3}\times S^{3}$ and
$AdS_{3}\times S^{3}$ algebraic curves arising from a giant magnon
$G(z)=G(z,\Xpm)$ of polarisation ``3''. The left column lists all
modes in $\ads3\times\s3\times\s3\times\s1$ including the four massless
modes. Empty cells indicate modes absent from the algebraic curve
description. \smallskip \protect \\
The right columns list the modes in the $AdS_{3}\times S^{3}$ case
\cite{David:2010yg}, which can be labelled by a pair of sheets $(i,j)$.
Boson in the same row are identified, but the vertical placement of
fermions should be taken as random. \label{tab:List-of-phase-shifts} }
\end{table}

Note that all of the phase shifts are independent of $\phi$, and
that all are proportional to either $G(z)=G(z,\Xpm)$ or $G(\tfrac{1}{z})$.
Let us define expansion coefficients of this function to be $\bar{Q}_{n}$
as follows:
\begin{equation}
G(z,\Xpm)\:=\:-i\log\frac{z-\Xp}{z-\Xm}-\frac{p}{2}\:=\:\sum_{n=0}^{\infty}-\bar{Q}_{n+1}z^{n}.\label{eq:defn-G-Qbar}
\end{equation}
Note however that these coefficients are not quite the standard ones
$Q_{n}$, which are defined for the resolvent $-i\log\frac{z-\Xp}{z-\Xm}$
alone. The term $-p/2$ is a twist added to the quasimomenta to allow
for the fact that one giant magnon is not a closed string \cite{Gromov:2008ie,Gromov:2007ky}.
Thus 
\begin{equation}
Q_{1}=p,\qquad\bar{Q}_{1}=\frac{p}{2}\label{eq:Q1-vs-Q1bar}
\end{equation}
where $p=-i\log\Xp/\Xm$, while for all higher charges 
\[
Q_{n+1}(\Xpm)=\bar{Q}_{n+1}(\Xpm)=\frac{i}{n}\Big(\frac{1}{\Xp^{n}}-\frac{1}{\Xm^{n}}\Big),\qquad n\geq1.
\]
At strong coupling these are 
\[
\bar{Q}_{n+1}(\Xpm)=\frac{2}{n}\sin\Big(n\frac{p_{X}}{2}\Big)+\mathcal{O}\Big(\frac{1}{h}\Big).
\]
This form can be used to obtain the following identity 
\begin{equation}
\sum_{\substack{r,s=1\\
r+s\:\mathrm{odd}
}
}^{\infty}\bar{Q}_{r}(p_{1})\bar{Q}_{s}(p_{2})=\frac{\pi^{2}}{2}\sign(p_{1}p_{2})+\mathcal{O}\Big(\frac{1}{h}\Big).\label{eq:closed-form-sum-QQ-new-1}
\end{equation}
derived using these sums: 
\begin{align*}
\smash{\sum_{\substack{n,m=1\\
n+m\:\mathrm{odd}
}
}^{\infty}}\frac{\sin(n\frac{p_{1}}{2})}{n}\;\frac{\sin(m\frac{p_{2}}{2})}{m} & =\frac{\pi}{16}\left(2\pi-\left|p_{1}\right|-\left|p_{2}\right|\right)\sign(p_{1}p_{2})\\
\sum_{n\geq1\:\mathrm{odd}}\frac{\sin(n\frac{p_{2}}{2})}{n} & =\frac{\pi}{4}\sign(p_{2}).
\end{align*}

Consider first the case of scattering two physical magnons of the
same polarisation ``3'', and the contribution of a mode for which
$\delta_{r}=\pm G(z)$. The integrals we need to do arise from sums
over $n$ using the method of \cite{SchaferNameki:2006gk}, which
uses the poles of $\cot(\pi n)$ to write a contour integral.%
\footnote{See for instance \cite{Abbott:2012dd} for a diagram of the contours.%
} This is then converted to an integral in $z$ with (\ref{eq:delta-is-two-pi-n-with-vac}),
and after deforming the contour to the unit circle $U$, we may approximate
$\cot(\pi n_{r}(z))=\pm i$ on the upper/lower half plane. The result
is this: 
\begin{equation}
\newI_{XY}=\sum_{\pm}\frac{\mp1}{16\pi}\Big\{\int_{U_{\pm}}+\int_{C_{\pm}}\Big\} dz\:\frac{\partial G(z,\Xpm)}{\partial z}G(z,\Ypm)\label{eq:integral-with-Upm}
\end{equation}
Here the contours $C_{\pm}$ are components surrounding the poles
and cuts outside the unit circle. 
\begin{itemize}
\item For the integral $I_{XY}$, the antisymmetric term in (\ref{eq:my-IXY-IX-organisation})
is 
\[
\newI_{YX}-\newI_{XY}=\frac{1}{8\pi}\sum_{\substack{r,s\geq1\\
r+s\:\mathrm{odd}
}
}c_{r,s}^{\mathrm{BLMT}}\;\bar{Q}_{r}(\Xpm)\bar{Q}_{s}(\Ypm)
\]
where $c_{r,s}^{\mathrm{BLMT}}$ are precisely the antisymmetrised
coefficients of \cite{Beccaria:2012kb}:
\begin{equation}
c_{r,s}^{\mathrm{BLMT}}=2\frac{s-r}{r+s-2},\qquad r+s\mbox{ odd},\; r,s\geq1.\label{eq:coeff-BLMT}
\end{equation}
For the symmetric term we can use the identity (\ref{eq:closed-form-sum-QQ-new-1})
above to get this: 
\[
\newI_{XY}+\newI_{YX}=\frac{1}{4\pi}\sum_{\substack{r,s\geq1\\
r+s\:\mathrm{odd}
}
}\bar{Q}_{r}\bar{Q}_{s}=\frac{\pi}{8}\sign(p_{1}p_{2})
\]

\item The $I_{X}$ integrals are surface terms. Evaluating these we find
\[
\newI_{X}=\frac{1}{4\pi}\sum_{\substack{r,s\geq1\\
r+s\:\mathrm{odd}
}
}\bar{Q}_{r}(\Xpm)\bar{Q}_{s}(\Xpm)=\frac{\pi}{16},\qquad\newI_{Y}=\frac{\pi}{16}
\]
again using identity (\ref{eq:closed-form-sum-QQ-new-1}). Then in
the total $\Delta\Theta_{r}$, (\ref{eq:my-IXY-IX-organisation}),
the symmetric term $I_{XY}+I_{YX}$ will cancel with the surface terms
$I_{X}+I_{Y}$ provided the signs of $p_{1}$ and $p_{2}$ are different. 
\item Finally, the contributions from $C_{\pm}$ (namely poles at $\Xpm$
and log cuts ending at $\Ypm$) all cancel within each integral. 
\end{itemize}
Then we must consider modes with $\delta_{r}=G(1/z)$, but in fact
these gives exactly the same results. 

It remains only to count the number of modes, using table \ref{tab:List-of-phase-shifts},
allowing for pre-factors and counting fermions with $-1$. Define
$\eta$ by 
\begin{equation}
\eta=\underset{\llap{\mbox{\ensuremath{{\scriptstyle m=}}}}\sin^{2}\phi,}{\quad\underbrace{-\,2}}\;\underset{\cos^{2}\phi,}{\underbrace{+\,8-2}}\;\underset{1}{\underbrace{-\,2}}=\begin{cases}
2,\qquad & AdS_{3}\times S^{3}\times S^{3}\times S^{1}\\
4, & AdS_{3}\times S^{3}\times T^{4}
\end{cases}\label{eq:counting-eta}
\end{equation}
where we omit modes of mass $\sin^{2}\phi$ in the $T^{4}$ case.
This is the only point of difference between the $T^{4}$ and $S^{1}$
cases. Then the final result $\Delta\Theta_{3\,3}$ in the $S^{1}$
case, or $\Delta\Theta_{(\tilde{1}\tilde{4}),(\tilde{1}\tilde{4})}$
in the $T^{4}$ case, is\vspace{-2mm} 
\begin{align}
\Delta\Theta(\Xpm,\Ypm) & =\frac{\eta}{8\pi}\sum_{\substack{r,s\geq1\\
r+s\:\mathrm{odd}
}
}c_{r,s}^{\mathrm{BLMT}}\;\bar{Q}_{r}(\Xpm)\bar{Q}_{s}(\Ypm)\nonumber \\
 & =\frac{\eta}{2}\sum_{\substack{r,s\geq1\\
r+s\:\mathrm{odd}
}
}c_{r,s}^{\mathrm{BOSST}}\; Q_{r}(\Xpm)Q_{s}(\Ypm).\label{eq:dTheta-final}
\end{align}
On the second line we use (\ref{eq:Q1-vs-Q1bar}) to express the correction
in terms of $Q_{n}$ rather than $\bar{Q}_{n}$, and absorb a factor
$\tfrac{1}{4\pi}$. The resulting coefficients are 
\begin{align}
c_{r,s}^{\mathrm{BOSST}} & =\frac{1}{4\pi}\left[2\frac{s-r}{r+s-2}-\delta_{r,1}+\delta_{1,s}\right],\qquad r+s\mbox{ odd},\; r,s\geq1\label{eq:coeff-final}\\
 & =\begin{cases}
\frac{1}{8\pi}c_{r,s}^{\mathrm{BLMT}},\qquad & r=1\mbox{ or }s=1\\
\frac{1}{4\pi}c_{r,s}^{\mathrm{BLMT}}, & r,s\geq2.
\end{cases}\nonumber 
\end{align}
This is the correction to the phase of the S-matrix, which is conventionally
written as $S=\hat{S}\sigma^{2}$ with the dressing phase $\sigma=e^{i\theta}$.
Thus $\Delta\Theta=2\theta^{\mathrm{HL}}$, with which we see perfect
agreement with \cite{Borsato:2013hoa} for the $T^{4}$ case.

\subsection{The left-right phase\label{sub:The-left-right-phase}}

In the above derivation we can equally well start from a classical
scattering state of two different polarisations. When we scatter one
left particle and one right particle (for instance by taking $r_{1}=3$,
$r_{2}=\bar{3}$) then the correction is different, as was observed
in the $T^{4}$ case by \cite{Beccaria:2012kb}.%
\footnote{See also \cite{Hoare:2011fj} for an earlier appearance of two independent
dressing phases. %
} 

The only integral we need consider is the case when $\delta_{rX}(z)=G(z,\Xpm)$
and $\delta_{rY}(z)=-G(\tfrac{1}{z},\Ypm)$. For this, 
\begin{align*}
\newI_{YX}-\newI_{XY} & =\frac{1}{8\pi}\sum_{\substack{r,s\geq1\\
r+s\:\mathrm{odd}
}
}\bar{c}_{r,s}^{\mathrm{BLMT}}\;\bar{Q}_{r}(\Xpm)\bar{Q}_{s}(\Ypm)\\
\shortintertext{where}\bar{c}_{r,s}^{\mathrm{BLMT}} & =-2\frac{r+s-2}{s-r},\qquad r+s\mbox{ odd},\; r,s\geq1.
\end{align*}
Then adding up all polarisations we get the same factor $\eta$ (\ref{eq:counting-eta}),
and writing the result in terms of the $Q_{n}$ we again see agreement
with \cite{Borsato:2013hoa} in the $T^{4}$ case:%
\footnote{Strictly this is $\Delta\Theta_{(\tilde{1}\tilde{4}),(\tilde{2}\tilde{3})}$
in the $T^{4}$ case.%
} 
\begin{align}
\Delta\Theta_{3\,\bar{3}}(\Xpm,\Ypm) & =\frac{\eta}{2}\sum_{\substack{r,s\geq1\\
r+s\:\mathrm{odd}
}
}\tilde{c}_{r,s}\; Q_{r}(\Xpm)Q_{s}(\Ypm)\label{eq:final-LR-phase}\\
\shortintertext{where}\tilde{c}_{r,s}^{\mathrm{BOSST}} & =\frac{1}{4\pi}\left[-2\frac{r+s-2}{s-r}+\delta_{r,1}-\delta_{1,s}\right],\qquad r+s\mbox{ odd},\; r,s\geq1.\label{eq:coeff-LR}
\end{align}

The relation between these two phases and the original $AdS_{5}$
HL phase is this: 
\begin{equation}
c_{r,s}^{\mathrm{BOSST}}+\tilde{c}_{r,s}^{\mathrm{BOSST}}=\frac{1}{4\pi}\left[c_{r,s}^{\mathrm{BLMT}}+\bar{c}_{r,s}^{\mathrm{BLMT}}\right]=\frac{1}{4\pi}\left[-8\frac{(r-1)(s-1)}{(r+s-2)(s-r)}\right]\qquad\label{eq:c-plus-ctilde-equals-HL}
\end{equation}
The factor in square brackets is precisely $c_{r,s}$ defined in \cite{Hernandez:2006tk}.
Note that the terms with $r=1$ or $s=1$ cancel here, and that the
modifications of \cite{Borsato:2013hoa} likewise cancel.

\subsection{Particles of different mass%
\footnote{I thank the authors of \cite{Borsato:2013hoa} for suggesting this
clarification, expanded from a single sentence in v1. %
}\label{sub:Particles-of-unequal-mass}}

In the $S^{1}$ case there is, in addition to a division into left-
and right-sector particles, a distinction between light particles
of mass $\sin^{2}\phi$, mass $\cos^{2}\phi$, and heavy particles
of mass $1$. The one-loop scattering phase for two light particles
of different masses is the same, while the heavy modes behave as composites
of two light particles. 

To show this we must compute $\Delta\Theta_{ab}$ for all $a,b$,
and it is sufficient to show that the following factor does not vary:
\begin{equation}
\sum_{r}(-1)^{F_{r}}\:\partial_{z}\delta_{ra}(z)\:\delta_{rb}(z)=2\big[G'(z)\, G(z)-\partial_{z}G(\tfrac{1}{z})\, G(\tfrac{1}{z})\big].\label{eq:sum-delta-delta}
\end{equation}
The phase shifts $\delta_{ra}$ needed for light bosons are trivially
related to $\delta_{r3}$ as used above; they are shown in table \ref{tab:Phase-shifts-various}. 

The one-loop phase is also the same if one of the particles is a fermion:
$\Delta\Theta_{3,1f}=\Delta\Theta_{33}$. In this case the classical
soliton needed for $\delta_{r,1f}(z,\Xpm)$ is no longer a cromulent
giant magnon, but proceeding all the same by turning on $G_{1}=G_{2}=\smash{\tfrac{1}{2\sin^{2}\phi}}G(z)$
(i.e. making a giant $1f$ mode, see table \ref{tab:List-of-modes})
we get the following quasimomenta: 
\begin{align}
p_{1}(x) & =\:\frac{1}{2\sin^{2}\phi}G(x,\Xpm)\nonumber \\
p_{2}(x) & =\frac{\Delta}{2g}\,\frac{x}{x^{2}-1}+\frac{1}{2\sin^{2}\phi}G(x,\Xpm)\label{eq:quasi123-fermi1f}\\
p_{3}(x) & =0.\nonumber 
\end{align}
This and the similar giant $3f$ mode with $G_{3}=G_{2}=\smash{\tfrac{1}{2\cos^{2}\phi}}G(z)$
lead to the phase shifts on the right of table \ref{tab:Phase-shifts-various}.
In this case they are not independent of $\phi$, but nevertheless
the total (\ref{eq:sum-delta-delta}) is the same, and hence $\Delta\Theta$
is too. 

Finally we can also consider scattering with one of the heavy modes.
These are not fundamental particles in the Bethe equations, but in
the worldsheet theory they are simply less strongly curved directions
in space. The heavy bosons are modes in $AdS_{3}$ directions, and
thus not giant magnons. From the integrable point of view they are
a sum of two fermions, $4=1f+3f$, and hence the relevant phase shifts
are 
\[
\delta_{r,\,4}(z,\Xpm)=\delta_{r,\,1f}(z,\Xpm)+\delta_{r,\,3f}(z,\Xpm).
\]
Then the result for heavy-light scattering is clearly $\Delta\Theta_{34}=2\Delta\Theta_{33}$,
and for heavy-heavy $\Delta\Theta_{44}=4\Delta\Theta_{33}$.

\begin{table}
\small \hspace{-10mm}
\begin{tabular}{c|cc|cc|cc|cc|}
$\negthickspace\negthickspace r$ & \multicolumn{2}{c|}{$\delta_{r,\,\bar{3}\,}(z)$} & \multicolumn{2}{c|}{$\delta_{r,\,1\,}(z)$} & \multicolumn{2}{c|}{$\delta_{r,\,1f}(z)$} & \multicolumn{2}{c|}{$\delta_{r,\,3f}(z)$}\tabularnewline
\hline 
$\negthickspace\negthickspace1$, $\bar{1}$ & $0$, & $\negthickspace\negthickspace0$ & $2G(z)$, & $\negthickspace\negthickspace-2G(\tfrac{1}{z})$ & $G(x)$, & $-G(\tfrac{1}{z})$ & $-\tan^{2}\phi\: G(z)$, & $\tan^{2}\phi\: G(\tfrac{1}{z})$\tabularnewline
$\negthickspace\negthickspace1f$, $\bar{1}f$ & $G(\tfrac{1}{z})$, & $\negthickspace\negthickspace-G(z)$ & $G(z)$, & $\negthickspace\negthickspace-G(\tfrac{1}{z})$ & $0$, & $0$ & $-\sec^{2}\phi\: G(z)$, & $\sec^{2}\phi\: G(\tfrac{1}{z})$\tabularnewline
\hline 
$\negthickspace\negthickspace3$, $\bar{3}$ & $-2G(\tfrac{1}{z})$, & $\negthickspace\negthickspace2G(z)$ & $0$, & $\negthickspace\negthickspace0$ & $-\cot^{2}\phi\: G(z)$, & $\cot^{2}\phi\: G(\tfrac{1}{z})$ & $G(x)$, & $-G(\tfrac{1}{z})$\tabularnewline
$\negthickspace\negthickspace3f$, $\bar{3}f$ & $-G(\tfrac{1}{z})$, & $\negthickspace\negthickspace G(z)$ & $-G(z)$, & $\negthickspace\negthickspace G(\tfrac{1}{z})$ & $-\cosec^{2}\phi\: G(z)$, & $\negthickspace\cosec^{2}\phi\: G(\tfrac{1}{z})$ & $0$, & $0$\tabularnewline
\hline 
$\negthickspace\negthickspace4$, $\bar{4}$ & $0$, & $\negthickspace\negthickspace0$ & $0$, & $\negthickspace\negthickspace0$ & $-\cosec^{2}\phi\: G(z)$, & $\negthickspace\cosec^{2}\phi\: G(\tfrac{1}{z})$ & $-\sec^{2}\phi\: G(z)$, & $\sec^{2}\phi\: G(\tfrac{1}{z})$\tabularnewline
$\negthickspace\negthickspace4f$, $\bar{4}f$ & $-G(\tfrac{1}{z})$, & $\negthickspace\negthickspace G(z)$ & $G(z)$, & $\negthickspace\negthickspace-G(\tfrac{1}{z})$ & $-\cot^{2}\phi\: G(z)$, & $\cot^{2}\phi\: G(\tfrac{1}{z})$ & $-\tan^{2}\phi\: G(z)$, & $\tan^{2}\phi\: G(\tfrac{1}{z})$\tabularnewline
\hline 
\end{tabular}\caption{Phase shifts of mode $r$ against various magnons. The first two columns
$\bar{3}$ (for left-right scattering) and $1$ (for unequal-mass)
are just re-arrangements of table \ref{tab:List-of-phase-shifts}.
The last two columns show phase shifts against fermions.  \label{tab:Phase-shifts-various}}
\end{table}

Note again that this discussion is limited to the $S^{1}$ case. In
the $T^{4}$ case, the scattering of two AdS modes $(\hat{1},\hat{4})$
has exactly the same $\Delta\Theta$ as two sphere modes $(\tilde{1},\tilde{4})$.
If we are to regard the $T^{4}$ case as the limit $\phi\to0$, then
this is an additional discontinuity, and in the opposite direction
to (\ref{eq:counting-eta}) above. These two discontinuities can be
summarised as follows: 
\begin{align}
\text{Sphere:}\qquad\Delta\Theta_{(\tilde{1}\tilde{4}),(\tilde{1}\tilde{4})} & =2\:\Delta\Theta_{3,3}\nonumber \\
\text{AdS:}\qquad\Delta\Theta_{(\hat{1}\hat{4}),(\hat{1}\hat{4})} & =2\:\Delta\Theta_{3,3}=\frac{1}{2}\,\Delta\Theta_{4,4}.\label{eq:both-steps-factor-2}
\end{align}

\subsection{Dependence on cutoff prescription%
\footnote{This section added in v2. The original preprint mistakenly claimed
that this term cancelled out.%
}\label{sub:The-c-term}}

So far we have ignored the last term in (\ref{eq:DeltaTheta-sum-n}),
with $\Delta E'\,\Theta_{\mathrm{cl.}}$. While there is no explicit
integral, this term will depend on the cutoff prescription used through
$\Delta E$, the one-loop term in the dispersion relation, which was
calculated using in (\ref{eq:dE-simple}) in \cite{Abbott:2012dd}.
This is  
\[
\Delta E=2c\sin\frac{p}{2}
\]
where $c=0$ when using a cutoff in the spectral plane, and is given
by (\ref{eq:rel-h-g}) when using a cutoff on the physical energy.
Then the term becomes 
\begin{equation}
-\frac{\Delta E_{1}'(p_{1})-\Delta E_{2}'(p_{2})}{g\Eclassone'(p_{1})-g\Eclasstwo'(p_{2})}\: g\Theta_{\mathrm{cl.}12}(p_{1},p_{2})\;=\:-\frac{c}{2}\Theta_{\mathrm{cl.}12}+\bigodiv{g}.\label{eq:last-term-dE-with-c}
\end{equation}

This constant $c$ is the one-loop term in the relation between the
Bethe coupling and the string tension: $h=2g+c+\bigo{1/g}$. The former
is the coupling appearing in the integrable structure, while the latter
appears in any pure string calculation. Based on what we learned in
$AdS_{4}\times CP^{3}$ \cite{McLoughlin:2008he,Gromov:2008fy,Bandres:2009kw,Abbott:2010yb,Abbott:2011xp}
we might expect the last term in (\ref{eq:DeltaTheta-sum-n}) to arise
as a result of expanding in $g$ some exact result depending on $h$
but not on $c$. Comparing the expansion in $g$ from (\ref{eq:CDLM-notation-g})
above to one in $h$, we do indeed find such a term: 
\begin{equation}
\Theta\;=\; g\Theta_{\mathrm{cl.}}+\Delta\Theta+\bigodiv{g}\;=\;\frac{h}{2}\Theta_{\mathrm{cl.}}+\left[\Delta\Theta-\frac{c}{2}\Theta_{\mathrm{cl.}}\right]+\bigodiv{h}.\label{eq:expanding-Theta}
\end{equation}
However the sign of (\ref{eq:last-term-dE-with-c}) is not right for
this explanation --- this term is part of $\Delta\Theta$ and so doubles
the $-\frac{c}{2}\Theta_{\mathrm{cl.}}$ in square brackets, rather
than canceling it. Thus using a different cutoff (and hence a different
$c$) will lead us to a different scattering phase $\Theta(h)$. 

Something conceptually similar was seen in \cite{Abbott:2012dd},
where changing the cutoff prescription led to a change in the one-loop
term in the cusp anomalous dimension $f(h)$. It was argued there
that this meant the cutoff was not a matter of indifference, and that
there must be a preferred choice. And further that if we demand a
smooth $\phi\to0$ limit, then the preferred choice was the physical
cutoff. Here we saw that a smooth connection to the $T^{4}$ case
seems to be ruled out, (\ref{eq:both-steps-factor-2}), undercutting
this argument. So the question of what prescription is preferred may
still be open. 

\subsection{A very small $p$ limit\label{sub:A-very-small-p-limit}}

The results above assume $p\sim1$ and strong coupling, particularly
in (\ref{eq:closed-form-sum-QQ-new-1}). This section looks very briefly
at the result of expanding instead at small $p$, small enough that
we may approximate 
\begin{equation}
Q_{n}(\Xpm)=(p_{X})^{n}\Big(\frac{h}{2m}\Big)^{n-1}+\bigo{p_{X}^{\: n+2}h^{n-1}(1+h^{2})}.\label{eq:very-small-p}
\end{equation}
It is a novel feature of the $AdS_{3}$ dressing phase that $\Delta\Theta$
has terms $c_{1,s}Q_{1}Q_{s}$ which can contribute at linear order
in this limit. In terms of the integrals above, the antisymmetric
term gives 
\begin{equation}
\newI_{YX}-\newI_{XY}=\frac{1}{4\pi}\sum_{s\geq2\:\mathrm{even}}Q_{s}(\Ypm)p_{X}+\bigo{p_{X}}^{2}.\label{eq:very-small-p-linear-term}
\end{equation}

However, notice that the symmetric term gives exactly the same result:
\[
\newI_{XY}+\newI_{YX}=\frac{1}{4\pi}\sum_{s\geq2\:\mathrm{even}}Q_{s}(\Ypm)p_{X}+\bigo{p_{X}}^{2}.
\]
This is no longer cancelled by the surface terms, as the first term
in $I_{X}$ is $Q_{1}Q_{2}\sim p_{X}^{\,3}$, and clearly $I_{Y}$
is free of $p_{X}$. But it comes with a pre-factor 
\[
\frac{\Eclass'(p_{X})+\Eclass'(p_{Y})}{\Eclass'(p_{X})-\Eclass'(p_{Y})}=-1+\bigo{p_{X}}
\]
and thus cancels the antisymmetric term. 

\section{Summary and Conclusions\label{sec:Conclusions}}

This note gives the first direct semiclassical calculation of the
one-loop dressing phase for strings in $AdS_{3}\times S^{3}\times S^{3}\times S^{1}$.
The result, when expanded in the charges $Q_{n}$, is precisely one
half the coefficients found by BOSST \cite{Borsato:2013hoa} from
their all-loop dressing phase for $AdS_{3}\times S^{3}\times T^{4}$.
If we sum over only those virtual modes which remain massive in the
$T^{4}$ limit, then this matches \cite{Borsato:2013hoa} exactly. 

Both the $S^{1}$ and $T^{4}$ cases feature a division of particles
into left and right sectors, and the HL phase for the scattering of
two particles of the same persuasion is different to that for the
scattering of one of each. In the $S^{1}$ case there is also a division
between particles of mass $\sin^{2}\phi$ and those of mass $\cos^{2}\phi$,
but the HL phase is indifferent to this.%
\footnote{The full theory contains also modes of mass $1$ (which are composite
in the Bethe equation description) and massless modes (absent both
here and in the Bethe equations). %
}

Some comments on the calculation:
\begin{itemize}
\item Antisymmetry is built into this way of calculating the phase: if we
do not assume that $\Delta\Theta_{21}=-\Delta\Theta_{12}$ in (\ref{eq:result-of-path-1-1})
then equating it to (\ref{eq:result-of-path-2-1}) tells us nothing.
To get something like (\ref{eq:dTheta-one-sided-CDLM}) we must drop
some surface terms which vanish in $AdS_{5}$ but not here. 
\item In the analogous $AdS_{5}$ and $AdS_{4}$ calculations \cite{Gromov:2007cd,Chen:2007vs,Gromov:2008qe},
all modes with classical phase shift $G(z)$ or $G(\tfrac{1}{z})$
cancel out, and the entire phase is given by modes $\delta_{r}(z)=G(z)-G(\tfrac{1}{z})$.
(See table \ref{tab:List-of-phase-shifts-AdS5}.) This clearly vanishes
at $z=\pm1$, which is what removes the surface terms. 
\item The integral implementing the infinite sum over all modes $n$ is
finite for each polarisation considered (and the integrand regular
at $z=\pm1$) so should not be sensitive to the cutoff used. This
is in contrast to the case of energy corrections (\ref{eq:dE-simple}),
where the contribution of one boson or fermion diverges quadratically.%
\footnote{For discussion of this see \cite{Gromov:2008fy,McLoughlin:2008he,Bandres:2009kw,Abbott:2010yb,Abbott:2011xp}
in $AdS_{4}$ and \cite{Sundin:2012gc,Abbott:2012dd,Beccaria:2012kb}
in $AdS_{3}$. %
} However one term in $\Delta\Theta$ arises from the magnons' one-loop
energy corrections $\Delta E_{i}$, and through this depends on the
cutoff-dependent subleading term in $h=2g+c+\bigo{1/g}$. The one-loop
term when expanding in $h$ is not independent of $c$. This is discussed
in section \ref{sub:The-c-term}, and elsewhere assumed to vanish.
\item The phase for the $S^{1}$ case is independent of $\phi$ (or $\alpha$).
However to see agreement with the $T^{4}$ case we must omit those
virtual modes which become massless in this limit, (\ref{eq:counting-eta}).
Thus if the $T^{4}$ case is to be thought of as a limit $\phi\to0$
then it is a discontinuous one. Some related discontinuities were
seen in worldsheet calculations in \cite{Beccaria:2012kb}. Note also
that the scattering of two heavy modes has $4$ times the $\Delta\Theta$,
but in the $T^{4}$ case this distinction is lost. This is a second
discontinuity, and in the opposite direction, see (\ref{eq:both-steps-factor-2}). 
\item When working in the $T^{4}$ case (and apart from the issue of antisymmetrisation)
the difference between this calculation and that of \cite{Beccaria:2012kb}
is equation (\ref{eq:Q1-vs-Q1bar}), arising from the twists needed
in the quasimomenta \cite{Gromov:2007ky,Gromov:2008ie}. These were
treated correctly in \cite{David:2010yg}'s calculation of the HL
phase, although this for the left-left case only, and the result was
not expanded in terms of the charges $Q_{n}$. 
\item Finally, and more speculatively, note that for the linear term when
expanding $\Delta\Theta$ at small $p$, which is proportional to
$c_{1,s}$ (and thus absent in $AdS_{5}$), there is a cancellation
between the antisymmetrised term and the symmetric term $I_{XY}+I_{YX}$
(which in the magnon regime instead cancels surface terms). 
\end{itemize}
Some comments on relations to other work:

The near-flat-space limit calculation of \cite{Sundin:2013ypa} found
agreement with the phase given by \cite{Beccaria:2012kb}.%
\footnote{The near-BMN theory derived there is for the $S^{1}$ case, with the
limit $\phi\to0$ taken before comparing with \cite{Beccaria:2012kb}.
It is not entirely clear whether a factor of 2 discrepancy should
be seen. %
} The lowest terms $c_{r,1}$ and $c_{1,s}$ are not visible in this
limit (the sum of all such is $h^{-1/2}$ compared to other terms
at $h^{0}$) and thus nothing changes when using the phase of \cite{Borsato:2013hoa}.
In the near-BMN limit all terms contribute at the same order, and
thus it would be very interesting if a way to avoid the divergences
seen by \cite{Sundin:2013ypa} could be found. 

Nothing in the derivation above makes any reference to the Bethe equations.
Of course in $AdS_{5}\times S^{5}$ the dressing phase was invented
to match the string's Bethe equations to those for the SYM spin chain
\cite{Minahan:2002ve,Arutyunov:2004vx}. And the goal of \cite{Beisert:2005cw,Hernandez:2006tk}
was to reproduce the energy of certain strings at one loop. 

Some comparisons between Bethe equations and $AdS_{3}$ string calculations
were explored in \cite{Beccaria:2012kb,Beccaria:2012pm}, using the
equations proposed in 2011 \cite{OhlssonSax:2011ms} which reduce
(in the sectors considered) to exactly the same $su(2)$ and $sl(2)$
Bethe equations used for $AdS_{5}\times S^{5}$. When calculating
the effect of the HL phase on the energy of a given string, the only
change needed in \cite{Hernandez:2006tk} is to start the sums at
$r,s=1$. In general the agreement was good but the interpretation
of some extra terms is not entirely clear. 

These comparisons will certainly be changed by the use of the coefficients
of \cite{Borsato:2013hoa}, and they should also be updated to test
the Bethe equations proposed in \cite{Borsato:2013qpa}. In the $S^{1}$
case we should use the coefficients derived here, and the relevant
Bethe equations are those of \cite{Borsato:2012ss}. I hope to return
to this topic in the near future.

\subsection*{Acknowledgements}

I am grateful to Justin David, Olof Ohlsson Sax, Valentina Puletti,
Per Sundin and Kostya Zarembo for various conversations, and to CERN
and CKA for hospitality. This work was supported by a UCT URC Postdoctoral
Research Fellowship. \phantom{ 
\cite{Borsato:2012ud} 
\cite{Ahn:2012hw, Sax:2012jv} 
} 
\phantom{ \smash{
\includegraphics[width=3cm]{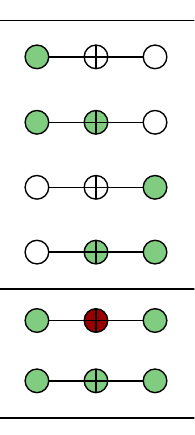}
Modes of the $AdS_3\times S^3\times S^3$ Algebraic Curve. \label{tab:List-of-modes} 
} }
\phantom{ \smash{
\includegraphics[width=3cm]{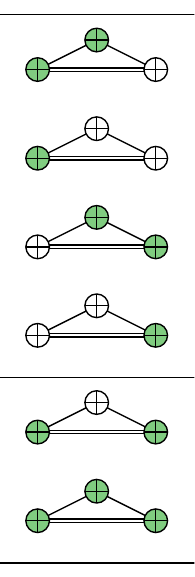}
Right-hand modes when using the mixed grading Cartan matrix $A'$, (\ref{eq:cartan-A-mixed}). \label{tab:List-of-modes-alt} 
} }

\appendix

\section{Algebraic Curve Setup\label{sec:Algebraic-Curve-Setup}}

Here for completeness we recall the parts of \cite{Abbott:2012dd}
we will need; this in turn was largely following \cite{Babichenko:2009dk,Zarembo:2010yz}.
Afterwards we will modify this to use the mixed grading suggested
by \cite{Borsato:2012ss} but the resulting system is equivalent. 

\begin{table}
\centering 

\begin{tabular}{cc|ccc|c}
 & $r$ & $\delta_{r}(z)$ &  & $r$ & $\delta_{r}(z)$\tabularnewline
\hline 
Bosons: & $(1,3)$ & $G(z)-G(\tfrac{1}{z})$ & \quad{}Fermions: & $(1,\hat{3}),(\hat{2},4)$ & $-G(\tfrac{1}{z})$\tabularnewline
 & $(1,4)$ & $-2G(\tfrac{1}{z})$ &  & $(1,\hat{4}),(\hat{1},4)$ & $-G(\tfrac{1}{z})$\tabularnewline
 & \textbf{$(2,3)$} & \textbf{$2G(z)$} &  & $(2,\hat{3}),(\hat{2},3)$ & $G(z)$\tabularnewline
 & $(2,4)$ & $G(z)-G(\tfrac{1}{z})$ &  & $(2,\hat{4}),(\hat{1},3)$ & $G(z)$\tabularnewline
Four $AdS$ bosons\hspace*{-1cm} &  & 0 &  &  & \tabularnewline
\hline 
\end{tabular}

\caption{Phase shifts in $AdS_{5}\times S^{5}$, for a background giant magnon
of $(2,3)$ polarisation. In fact $(1,3)$ and $(2,4)$ alone are
sufficient to derive the HL phase, the others cancel. These are precisely
the bosons absent from $AdS_{3}\times S^{3}$ (table \ref{tab:List-of-phase-shifts}).
\label{tab:List-of-phase-shifts-AdS5} }
\end{table}

The Cartan matrix for $d(2,1;\alpha)^{2}$ is 
\begin{equation}
A=\left[\begin{array}{ccc}
4\sin^{2}\phi & -2\sin^{2}\phi & 0\\
-2\sin^{2}\phi & 0 & -2\cos^{2}\phi\\
0 & -2\cos^{2}\phi & 4\cos^{2}\phi
\end{array}\right]\otimes1_{2\times2}.\label{eq:cartan-A}
\end{equation}
For each Cartan generator $\Lambda_{\ell}$ there is a quasimomentum
$p_{\ell}(x)$, where $\ell=1,2,3,\bar{1},\bar{2},\bar{3}$. In addition
to $A$, we also need to know the matrix $S$ which gives the inversion
symmetry: 
\begin{equation}
p_{\ell}(\tfrac{1}{x})=S_{\ell m}p_{m}(x),\qquad S=1_{3\times3}\otimes\left[\begin{array}{cc}
0 & 1\\
1 & 0
\end{array}\right].\label{eq:inv-cond-p}
\end{equation}
The vacuum algebraic curve has poles at $x=\pm1$, controlled by
a vector $\kappa_{\ell}$: 
\begin{equation}
p_{\ell}(x)=\frac{\kappa_{\ell}x}{x^{2}-1},\qquad\kappa=\frac{\Delta}{2g}(0,1,0,\:0,-1,0).\label{eq:vacuum-p}
\end{equation}

Solutions above this vacuum constructed by introducing various cuts.
The crucial equation is that when crossing a cut $C$ (of mode number
$n$) in sheet $\ell$, the change in $p_{\ell}$ is given by 
\begin{equation}
p_{\ell}\to p_{\ell}-A_{\ell m}p_{m}+2\pi n.\label{eq:jump-condition-p}
\end{equation}
By continuity this change must be zero at the end of the cut, and
this gives us an equation relating the position of the branch point
to the cut's mode number. The various polarisations (including heavy
modes) are summarised in table \ref{tab:List-of-modes}. 

Vibrational modes are very short cuts which may be treated as poles.
The resulting perturbation of the quasimomenta nearby is given by
\[
\delta p_{\ell}(x)=\frac{k_{\ell r}\alpha(y)}{x-y}+\bigo{x-y}^{0},\qquad\alpha(y)=\frac{1}{2g}\frac{y^{2}}{y^{2}-1}.
\]
The energy of this perturbation can be read off from $\delta p_{\ell}(x)$
near to $x=\infty$, and gives the off-shell frequency $\Omega(y)=\delta\Delta$.
The frequency of mode number $n$ is then given by evaluating this
at the point fixed by (\ref{eq:jump-condition-p}).

\newcommand{\smalldynkin}[3]{
\begin{tikzpicture}[scale=0.6]
\draw (1,0) -- (3,0);
\draw [fill=#1] (1,0) circle (2mm); 
\draw [fill=#2] (2,0) circle (2mm); 
\draw [fill=#3] (3,0) circle (2mm); 
\draw [thin] (18mm,0) -- (22mm,0);
\draw [thin] (2,-2mm) -- (2,2mm);
\end{tikzpicture}
} 

\newcommand{\colourone}{midgreen} \newcommand{\colourtwo}{darkred} 

\newcommand{\smallcirc}[1]{\begin{tikzpicture}[scale=0.6] \draw [fill=#1] (0,0) circle (2mm); \end{tikzpicture}}

\begin{table}
\centering{}%
\begin{tabular}{cccccc}
 & $r$ & $m_{r}$ &  & $\qquad2\pi n_{r}=-k_{\ell r}A_{\ell k}p_{k}\negthickspace\negthickspace\negthickspace\negthickspace\negthickspace\negthickspace\negthickspace\negthickspace\negthickspace\negthickspace\negthickspace\negthickspace$ & \tabularnewline
\hline 
$\phi^{1},\phi^{\bar{1}}$ & $1$, $\bar{1}$ & $\sin^{2}\phi$ & \smalldynkin{\colourone}{white}{white} & $A_{1k}p_{k}$  &  $-A_{\bar{1}k}p_{k}$\tabularnewline
$\psi^{1},\psi^{\bar{1}}$ & $1f$, $\bar{1}f$ & $\sin^{2}\phi$ & \smalldynkin{\colourone}{\colourone}{white} & $(A_{1k}+A_{2k})p_{k}$  & $(-A_{\bar{1}k}-A_{\bar{2}k})p_{k}$ \tabularnewline
$\phi^{3},\phi^{\bar{3}}$ & $3$, $\bar{3}$ & $\cos^{2}\phi$ & \smalldynkin{white}{white}{\colourone} & $A_{3k}p_{k}$ & $-A_{\bar{3}k}p_{k}$\tabularnewline
$\psi^{3},\psi^{\bar{3}}$ & $3f$, $\bar{3}f$ & $\cos^{2}\phi$ & \smalldynkin{white}{\colourone}{\colourone} & $(A_{2k}+A_{3k})p_{k}$ & $(-A_{\bar{2}k}-A_{\bar{3}k})p_{k}$\tabularnewline
\hline 
 & $4$, $\bar{4}$ & 1 & \smalldynkin{\colourone}{\colourtwo}{\colourone} & $(A_{1k}+2A_{2k}+A_{3k})p_{k}$ & $(-A_{\bar{1}k}-2A_{\bar{2}k}-A_{\bar{3}k})p_{k}$\tabularnewline
 & $4f$, $\bar{4}f$ & 1 & \smalldynkin{\colourone}{\colourone}{\colourone} & $(A_{1k}+A_{2k}+A_{3k})p_{k}$ & $(-A_{\bar{1}k}-A_{\bar{2}k}-A_{\bar{3}k})p_{k}$\tabularnewline
\hline 
\end{tabular}\caption[Fake caption without tikz figures.]{List of modes in the $AdS_{3}\times S^{3}\times S^{3}$ algebraic
curve. The colouring of the nodes is $-k_{\ell r}$ with $\smallcirc{\colourone}= +1, -1$
and $\smallcirc{\colourtwo}= +2, -2$. The first column is the names
of the modes in \cite{Borsato:2012ud}.\label{tab:List-of-modes} }
\end{table}

The giant magnon is described by 
\begin{equation}
G(x,\Xpm)=-i\log\frac{x-\Xp}{x-\Xm}+\frac{i}{2}\log\frac{\Xp}{\Xm}.\label{eq:giant-magnon-G-with-twist}
\end{equation}
The second term here is a twist of the type first discussed in \cite{Gromov:2007ky},
and employed for giant magnons in \cite{Gromov:2008ie} where they
play an important role in finite-size corrections.%
\footnote{For similar L\"{u}scher F-term corrections in $AdS_{4}\times CP^{3}$,
it was necessary to introduce similar twists in \cite{Abbott:2011tp}
compared to the ansatz of \cite{Lukowski:2008eq}.%
} It is needed to take account of the fact that a giant magnon is not
a closed string; alternatively, it is a result in working in some
$\mathbb{Z}_{n}$ orbifold \cite{Astolfi:2007uz} in which the string
is closed. (See \cite{Ideguchi:2004wm,Beisert:2005he} for earlier
work on $\mathbb{Z}_{n}$ symmetries.)  The same twists were used
for the construction of BMN modes in \cite{Abbott:2012dd}. 

Choosing to consider a giant ``3'' mode we are interested in the
curve 
\begin{align}
p_{1}(x) & =0\nonumber \\
p_{2}(x) & =\frac{\Delta}{2g}\,\frac{x}{x^{2}-1}\label{eq:magnon-3-p}\\
p_{3}(x) & =\frac{1}{2\cos^{2}\phi}G(x,\Xpm).\nonumber 
\end{align}
This has momentum $p=-i\log(\Xp/\Xm)$, charge $Q=\smash{\frac{-ih}{2\cos^{2}\phi}}(\Xp+1/\Xp-\mbox{c.c.})=1$
for an elementary magnon, and dispersion relation 
\begin{equation}
E(p)=\Delta-J'=-i\frac{h}{2}\Big(\Xp-\frac{1}{\Xp}-\Xm+\frac{1}{\Xm}\Big)=\sqrt{Q^{2}\cos^{2}\phi+4h^{2}\sin^{2}\frac{p}{2}}.\label{eq:magnon-disp-rel}
\end{equation}
The Bethe coupling%
\footnote{This coupling is identified with $\sqrt{\lambda}$ in $\ads5$, but
not in $\ads4$ \cite{Nishioka:2008gz,Gaiotto:2008cg,Grignani:2008is}
where it was further investigated at strong coupling by \cite{McLoughlin:2008he,Gromov:2008fy,Bandres:2009kw,Abbott:2010yb,Abbott:2011xp}
and at weak coupling by \cite{Minahan:2009aq,Minahan:2009wg,Leoni:2010tb}.
Perturbative checks at weak coupling have now been done up to six
loops \cite{Bak:2009tq}, and eight loops \cite{Mauri:2013vd} at
which the first influence of the dressing phase appears.%
} $h$ is related to $g=R^{2}/4\pi\alpha'^{2}=\sqrt{\lambda}/4\pi$
(half the effective string tension) by%
\footnote{Note that \cite{Beccaria:2012kb,Borsato:2013hoa} define $h$ differently,
and \cite{Sundin:2012gc} defines $g$ differently. My conventions
follow \cite{David:2010yg,OhlssonSax:2011ms,Borsato:2012ss}.%
} 
\begin{equation}
h=2g+\frac{\sin^{2}\phi\,\log(\sin^{2}\phi)+\cos^{2}\phi\,\log(\cos^{2}\phi)}{2\pi}+\mathcal{O}\Big(\frac{1}{g}\Big).\label{eq:rel-h-g}
\end{equation}
The subleading term was first written down by \cite{David:2010yg}
for $\phi=0$ and by \cite{Sundin:2012gc,Abbott:2012dd,Beccaria:2012kb}
for general $\phi$. This assumes we are using the ``physical''
prescription, i.e. using a cutoff on energy. Using a cutoff in the
spectral plane, the subleading term vanishes \cite{Abbott:2012dd}.

\subsection{Mixed grading}

\newcommand{\smalldynkinalt}[3]{
\begin{tikzpicture}[scale=0.6]

\draw [double] (1,0) -- (3,0);
\draw (1,0) -- (2,0.5) -- (3,0);

\draw [fill=#1] (1,0) circle (2mm); 
\draw [fill=#2] (2,0.5) circle (2mm); 
\draw [fill=#3] (3,0) circle (2mm); 

\draw [thin] (8mm,0) -- (12mm,0);
\draw [thin] (1,-2mm) -- (1,2mm);

\draw [thin] (18mm,0.5) -- (22mm,0.5);
\draw [thin] (2,3mm) -- (2,7mm);

\draw [thin] (28mm,0) -- (32mm,0);
\draw [thin] (3,-2mm) -- (3,2mm);

\end{tikzpicture}
} 
\newcommand{\smallcircplus}[1]{
\begin{tikzpicture}[scale=0.6] 
\draw [fill=#1] (0,0) circle (2mm); 
\draw [thin] (-2mm,0) -- (2mm,0);
\draw [thin] (0,-2mm) -- (0,2mm);
\end{tikzpicture}}%

According to \cite{Borsato:2012ss} we should use a $d(2,1;\alpha)$
Dynkin diagram with different grading for modes on the right. Drawing
$\smallcircplus{white}$ for a fermionic node (and with labels to
indicate momentum-carrying roots) the left and right diagrams are:
\[ 
\begin{tikzpicture} [scale=0.5, baseline, semithick]

\draw (1,0) -- (2,1.73205) -- (3,0);

\draw [fill=white] (1,0) circle (3mm) node [anchor=north, yshift=-2mm] {\small 1}; 
\draw [fill=white] (2,1.73205) circle (3mm); 
\draw [fill=white] (3,0) circle (3mm) node [anchor=north, yshift=-2mm] {\small 1}; 

\draw [thin] (17mm,1.73205) -- (23mm,1.73205);
\draw [thin] (2,1.43205) -- (2,2.03205);

\end{tikzpicture}
\hspace{1cm}
\begin{tikzpicture} [scale=0.5, baseline, semithick]

\draw [double] (1,0) -- (3,0);
\draw (1,0) -- (2,1.73205) -- (3,0);

\draw [fill=white] (1,0) circle (3mm) node [anchor=north, yshift=-2mm] {\small 1}; 
\draw [fill=white] (2,1.73205) circle (3mm); 
\draw [fill=white] (3,0) circle (3mm) node [anchor=north, yshift=-2mm] {\small 1}; 

\draw [thin] (7mm,0) -- (13mm,0);
\draw [thin] (1,-3mm) -- (1,3mm);

\draw [thin] (17mm,1.73205) -- (23mm,1.73205);
\draw [thin] (2,1.43205) -- (2,2.03205);

\draw [thin] (27mm,0) -- (33mm,0);
\draw [thin] (3,-3mm) -- (3,3mm);

\end{tikzpicture}
\]  Above we used the left-hand diagram for both left and right modes,
as in \cite{Abbott:2012dd} and \cite{OhlssonSax:2011ms,Zarembo:2010yz}.
In this appendix we explicitly change the set-up to use the mixed
grading, and confirm that (as expected) this does not affect anything. 

To do this we now use the following Cartan matrix: 
\begin{equation}
A'=\left[\begin{array}{ccc}
4\sin^{2}\phi & -2\sin^{2}\phi & 0\\
-2\sin^{2}\phi & 0 & -2\cos^{2}\phi\\
0 & -2\cos^{2}\phi & 4\cos^{2}\phi
\end{array}\right]\oplus\left[\begin{array}{ccc}
0 & 2\sin^{2}\phi & -2\\
2\sin^{2}\phi & 0 & 2\cos^{2}\phi\\
-2 & 2\cos^{2}\phi & 0
\end{array}\right].\label{eq:cartan-A-mixed}
\end{equation}
With this we need 
\begin{equation}
S=\left[\begin{array}{ccc}
1 & 0 & 0\\
1 & -1 & 1\\
0 & 0 & 1
\end{array}\right]\otimes\left[\begin{array}{cc}
0 & 1\\
1 & 0
\end{array}\right].\label{eq:inv-cond-mixed}
\end{equation}
and vacuum
\begin{equation}
\kappa=-\frac{\Delta}{2g}(0,1,0,\:0,1,0).\label{eq:vac-mixed}
\end{equation}
Then the mode corresponding to ``1'' \smalldynkin{\colourone}{white}{white}
via inversion symmetry is now \smalldynkinalt{\colourone}{\colourone}{white},
colouring in two fermionic nodes to make a boson. All the others are
listed in table \ref{tab:List-of-modes-alt}.

\begin{table}
\centering{}%
\begin{tabular}{cccc}
$r$ & $m_{r}$ &  & $2\pi n_{r}$\tabularnewline
\hline 
$\bar{1}$ & $\sin^{2}\phi$ & \smalldynkinalt{\colourone}{\colourone}{white} & $-(A'_{\bar{1}k}+A'_{\bar{2}k})p_{k}$\tabularnewline
$\bar{1}f$ & $\sin^{2}\phi$ & \smalldynkinalt{\colourone}{white}{white} & $-A'_{\bar{1}k}p_{k}$ \tabularnewline
$\bar{3}$ & $\cos^{2}\phi$ & \smalldynkinalt{white}{\colourone}{\colourone} & $-(A'_{\bar{2}k}+A'_{\bar{3}k})p_{k}$\tabularnewline
$\bar{3}f$ & $\cos^{2}\phi$ & \smalldynkinalt{white}{white}{\colourone} & $-A'_{\bar{3}k}p_{k}$\tabularnewline
\hline 
$\bar{4}$ & 1 & \smalldynkinalt{\colourone}{white}{\colourone} & $(-A'_{\bar{1}k}-A'_{\bar{3}k})p_{k}$\tabularnewline
$\bar{4}f$ & 1 & \smalldynkinalt{\colourone}{\colourone}{\colourone} & $(-A'_{\bar{1}k}-A'_{\bar{2}k}-A'_{\bar{3}k})p_{k}$\tabularnewline
\hline 
\end{tabular}\caption[Fake caption without tikz figures.]{Right-hand modes when using the mixed grading Cartan matrix $A'$,
(\ref{eq:cartan-A-mixed}). Filled nodes \smallcirc{\colourone} for
$\ell=\bar{1},\bar{2},\bar{3}$ here indicate $k_{\ell r}=1$.  \label{tab:List-of-modes-alt} }
\end{table}

The giant magnon in the ``3'' polarisation now has
\begin{align}
p_{\bar{1}}(x) & =0\nonumber \\
p_{\bar{2}}(x) & =\frac{\Delta}{2g}\,\frac{x}{x^{2}-1}+\frac{1}{2\cos^{2}\phi}G(\tfrac{1}{x},\Xpm)\label{eq:magnon-3-pbar}\\
p_{\bar{3}}(x) & =\hphantom{\frac{\Delta}{2g}\,\frac{x}{x^{2}-1}+}\frac{1}{2\cos^{2}\phi}G(\tfrac{1}{x},\Xpm).\nonumber 
\end{align}
It is easy to check that when using this the phases $\delta_{r}$
in table \ref{tab:List-of-phase-shifts} are unchanged. 

\bibliographystyle{my-JHEP-4}
\bibliography{/Users/me/Documents/Papers/complete-library-processed,complete-library-processed}

\end{document}